\documentclass[plb,showpacs,preprintnumbers,nofootinbib,amsmath]{revtex4}
\usepackage[dvips,final]{graphicx}
  \usepackage{amssymb}
   \usepackage{amsmath}
    \usepackage{epsfig} 
     \usepackage{bm}
      \usepackage{pifont}

\def\({\left(}
\def\[{\left[}
\def\){\right)}
\def\]{\right]}

\begin{document}
\preprint{\hbox{RUB-TPII-01/2011}}

\title{Evolution of transverse-momentum-dependent densities}
\author{I.~O.~Cherednikov\footnote{Also at:
{\sl ITPM, Moscow State Univerisity, Russia}}}
\email{igor.cherednikov@jinr.ru}
\affiliation{Departement Fysica, Universiteit Antwerpen, B-2020 Antwerpen, Belgium\\
             and
             \\
             Bogoliubov Laboratory of Theoretical Physics,
             JINR,
             RU-141980 Dubna, Russia}
\author{N.~G.~Stefanis}
\email{stefanis@tp2.ruhr-uni-bochum.de}
\affiliation{Institut f\"{u}r Theoretische Physik II,
             Ruhr-Universit\"{a}t Bochum,
             D-44780 Bochum, Germany\\}

\date{\today}

\begin{abstract}
We discuss different operator definitions of the transverse-momentum dependent (TMD) parton
densities from the point of view of their renormalization-group (RG) properties
and UV evolution. We also consider the structure of the gauge links (Wilson lines) in
these operator definitions and examine the role of the soft factor in
the factorization formula within the TMD approach to semi-inclusive
processes.

\end{abstract}


\maketitle

The candidates for the {\it golden} measurement at the EIC are the
spin-dependent Sivers function $f_{1T}^{\perp}$, as well as the
unpolarized quark distribution $f_{1}$.
The proposed {\it silver} candidates are the transversity, the
Boer-Mulders, and the Collins functions.
All these objects are transverse-momentum dependent (TMD) parton
densities (termed in the following for simplicity, TMDs) that
describe the inner structure of hadrons by taking into account the
partonic longitudinal {\it and also} the transverse degrees of freedom
\cite{Sop77, Col78, CS81, Col03, BR05}.

The QCD factorization formula for a semi-inclusive structure function
is expected to have the following symbolic form
(modulo power corrections)
\cite{CS81, CS82, JMY04}
\begin{equation}
  F
  \sim
 {H }  \otimes {{\cal F}_D}
  \otimes {{\cal F}_F} \otimes {S} \ .
  \label{eq:factor_symb}
\end{equation}
This expression contains the hard part $H$,
the distribution and fragmentation transverse-momentum dependent
functions ${{\cal F}_D}$ and ${{\cal F}_F}$, respectively, and the soft
part $S$, which is introduced in order to take care of light-cone
singularities that cannot be cured by regularization.
While the hard part $H$ can be evaluated order by order within the
framework of perturbative QCD, the TMDs are essentially nonperturbative
entities and have to be modeled or extracted from the data.
Thus, little can be said about these objects, except their general
operator definition and their evolution behavior with respect to the
energy and rapidity scales.
Contrary to the fully collinear, i.e., {\it integrated} PDFs, the
explicit operator definition of TMDs is quite problematic, owing to the
appearance of extra (light-cone) singularities, and remains under
active investigation (see, e.g.,
\cite{JY02, BJY02, BMP03, Col08, CS07, CS08, CS09, CRS07, Bacch08, SCK10}).

Let us concentrate on the following two
definitions\footnote{A generalized definition, which includes into the
Wilson lines the spin-dependent Pauli term
$F^{\mu\nu}[\gamma_\mu, \gamma_\nu]$, is worked out in Ref.\
\cite{CKS10}---see also \cite{SCK10}.} of an unpolarized quark TMD
\begin{eqnarray}
  &&   f_1 (x, \ \mathbf{k_\perp} | n)
=
  \frac{1}{2}
  \int \frac{d\xi^- d^2\bm{\xi}_\perp}{(2\pi)^3}
  {\rm e}^{-ik^{+}\xi^{-} + i \mathbf{k}_\perp
\cdot \bm{\xi}_\perp}
  \left\langle
              h |\bar \psi_i (\xi^-, \bm{\xi}_\perp)
              [\xi^-, \bm{\xi}_\perp;
   \infty^-, \bm{\xi}_\perp]_n^\dagger
\right. \nonumber \\
&& \left. \times
   [\infty^-, \bm{\xi}_\perp;
   \infty^-, \bm{\infty}_\perp]_\mathbf{l}^\dagger \
   \gamma^+ \ [\infty^-, \bm{\infty}_\perp;
   \infty^-, \mathbf{0}_\perp]_\mathbf{l}
   [\infty^-, \mathbf{0}_\perp; 0^-,\mathbf{0}_\perp]_n
   \psi_i (0^-,\mathbf{0}_\perp) | h
   \right\rangle
\label{eq:tmd_naive}
\end{eqnarray}
and its counter-part
\begin{equation}
  f_1 (x, \ \mathbf{k_\perp} | v)
  =
  f_1 (x, \ \mathbf{k_\perp} | n \to v)\ .
  \label{eq:tmd_v}
\end{equation}
following the so-called Amsterdam notations, e.g., \cite{TM95}, while
other TMDs can be analyzed within the same framework.
Gauge invariance is ensured by means of path-ordered contour-dependent
Wilson-line operators (gauge links) with the generic form
$
  { [y,x|\Gamma] }
=
  {\cal P} \exp
  \left[-ig\int_{x[\Gamma]}^{y}dz_{\mu} {\cal A}^\mu (z)
  \right]
$
where ${\cal A} \equiv t^a A^a$,
and one has to distinguish between longitudinal
$[ \ , \ ]_{[n,\, v]}$ and
transverse $[ \ , \ ]_{[\, \mathbf{l}\, ]}$
gauge links \cite{BJY02, BMP03}.
On the other hand, the definition used in lattice simulations contains,
in contrast to the (semi-)infinite Wilson lines in
Eqs.\ ({\ref{eq:tmd_naive}}) and ({\ref{eq:tmd_v}}),
the finite, i.e., the {\it direct} gauge link joining the two quark fields
\cite{Hae09, Hae09_1, Musch09}.

In the tree-approximation, the ``distribution of a quark in a quark''
is normalized as
$$
f_1^{\rm tree} (x, \ \mathbf{k_\perp} | n ) =
  \delta(1 - x ) \delta^{(2)} (\mathbf{k_\perp})
$$
and the integration over $\mathbf{k_\perp}$ yields formally the
usual collinear (integrated) PDF
$$
\int\! d^2 k_\perp\  f_1 (x, \ \mathbf{k_\perp} | n )
=
f_1^{\rm tree} (x) = \delta(1 - x ) \ .
$$

However, definition (\ref{eq:tmd_naive}), taken literally,
produces---beyond the tree-level---certain pathological divergences,
which belong to one of the following three classes:

\begin{enumerate}

  \item Usual {\it UV-singularities} $\sim \frac{1}{\varepsilon}$ from
  integrations over loop momenta, which can be removed by using the
  standard $R-$operation.

  \item Pure {\it rapidity divergences}, which only appear in the
  {\it unintegrated} case.
  They cancel in the integrated distributions, but they are present in
  the TMD case giving rise to logarithmic and double-logarithmic terms
  of the form $\sim \ln \zeta , \ \ln^2 \zeta$; they have to be
  resummed by a consistent procedure.

  \item {\it Overlapping divergences}, which contain both UV and
  soft singularities simultaneously:
  $
  \sim \frac{1}{\varepsilon}\ln \zeta ,
  $
  meaning that the UV pole $\varepsilon^{-1}$ mixes with a ``soft''
  divergence, regularized by the auxiliary parameter $\zeta$.

\end{enumerate}
In addition, one may encounter combinations of the above types of
singularities ensuing from the soft factor.
Apart from the UV divergences, all other singularities originate from
uncompensated light-cone artifacts, which stem either from the
lightlike gauge links (in covariant gauges), or from specific terms in
the gluon propagator in (singular) light-cone axial gauges.
While the singularities of the second class may be simply regularized
by a rapidity cutoff---which is ``separated'' from other variables---the
effect of the third class is more severe, because these
singularities affect the UV-renormalization procedure, change the
anomalous dimensions, and modify, therefore, the RG-evolution.

In order to avoid the above-mentioned problems, the following approaches
have been proposed in the literature:

\begin{enumerate}

\item Shift in covariant gauges the gauge links off the light-cone:
$ v^2 \neq 0$,  or use instead the non-lightlike axial
gauge $(v \cdot A) = 0$ \cite{CS81}.
This amounts to definition (\ref{eq:tmd_v}), but it may cause problems
in properly deriving factorization \cite{Col08}.

\item Stay on the light-cone, Eq. (\ref{eq:tmd_naive}), but subtract
some specific soft factor $R$, which is defined in such a way as to exactly
cancel the extra divergences \cite{CH00, CM04, Hau07}: Eq.\
(\ref{eq:tmd_naive}) is substituted by the ``subtracted''
function $ f_1 (n) \to f_1 (n) \cdot R^{-1}$.

\item Perform a direct regularization of the light-cone singularities
in the gluon propagator \cite{CFP80}
$
{1}/{q^+} \to {1}/{[q^+](\eta)} \ ,
$
where $\eta$ is an additional dimensional parameter \cite{CS07}.
In this case, a generalized renormalization is in order, which is
formally equivalent to multiplying the TMD PDF by a particular soft
factor \cite{KR87}.

\item Use the light-cone axial gauge, but supply it with the
Mandelstam-Leibbrandt (ML) pole prescription \cite{Man83, Lei84}:
$
{1}/{q^+} \ \to \  {1}/({q^+ + i0q^-})  \ \ \hbox{or equivalently} \
\  {q^-}/({q^+q^- + i0})
$.
Now the overlapping singularities do not appear from the outset---at
least at the level of the one-loop order---while the contribution of
the soft factor is reduced to unity, rendering the gauge-invariant
definition valid \cite{CS09}.

\end{enumerate}

Let us present the UV evolution equations for the above definitions.
The off-the-light-cone TMD (\ref{eq:tmd_v}) does not contain
{\it overlapping} singularities.
Therefore, the renormalization-group equation reads
\cite{JMY04, KMPLA89}
\begin{equation}
  \mu \frac{d }{d\mu} \ f_1 (x, \mathbf{k_\perp}, \mu | v )
  =
   \gamma_{\rm LC} \  f_1 (x, \mathbf{k_\perp}, \mu | v ) \  \ , \ \
   \gamma_{\rm LC} = \frac{3}{4} \ \frac{\alpha_s C_{\rm F}}{\pi}
   + O(\alpha_s^2) \ ,
\label{eq:uv_v}
\end{equation}
where $\gamma_{\rm LC}$ is the anomalous dimension of the bilocal quark
operator in the light-cone gauge.
If one factorizes out the soft contribution $R_v$, as proposed in
Ref.\ \cite{JMY04}, then the anomalous dimension changes and one has
\begin{equation}
 f_1 (v, \mu) \to f_1 ( v, \mu) \cdot R_v^{-1} \ \ , \ \
  \mu \frac{d}{d\mu} \ \left[ f_1 (v, \mu) \cdot R_v^{-1} \right]
  =
   ( \gamma_{\rm LC} - \gamma_{\rm R} ) \ f_1 (v, \mu)
   \cdot R_v^{-1} \ ,
\label{eq:uv_vr}
\end{equation}
where $\gamma_{\rm R}$ is the one-loop anomalous dimension of the soft
factor $R_v$.

In contrast, the anomalous dimension of the ``light-cone'' TMD, before
subtraction, deviates from $\gamma_{\rm LC}$ and this deviation is
determined by the cusp anomalous dimension \cite{CS07}:
\begin{equation}
  \mu \frac{d }{d\mu} \ f_1 (n)
  =
   ( \gamma_{\rm LC} - \gamma_{\rm cusp}  ) \  f_1 (n) \ .
\label{eq:uv_n}
\end{equation}
Hence, the generalized renormalization procedure restores the ``broken''
anomalous dimensions, so that one finds
\begin{equation}
  f_1 (n) \to f_1 (n) \cdot R^{-1} \ \ , \ \
  \mu \frac{d }{d\mu} \ \left[ f_1 (n) \cdot R_n^{-1} \right]
  =
    \gamma_{\rm LC} \  \left[ f_1 (n) \cdot R_n^{-1}  \right]\ .
\label{eq:uv_nr}
\end{equation}
In the light-cone gauge with the Mandelstam-Leibbrandt prescription,
Eq.\ (\ref{eq:tmd_naive}) supplemented by the soft factor $R$
yields ab initio an anomalous dimension without lightlike artifacts:
\begin{equation}
  \mu \frac{d }{d\mu} \ \left[ f_1 (n)^{\rm ML} \cdot R_n^{-1} \right]
  =
  \mu \frac{d }{d\mu} \  f_1 (n)^{\rm ML}
  =
    \gamma_{\rm LC} \  \left[ f_1 (n)^{\rm ML} \cdot R_n^{-1} \right]
  =
    \gamma_{\rm LC} \ f_1 (n)^{\rm ML}\ .
\label{eq:uv_ml}
\end{equation}
Thus, evolution equations can be established, while the evolution with
respect to the {\it rapidity} variable---either $\zeta$, or $\eta$
(depending on the approach applied)---constitutes a separate task.

To conclude, let us sketch a couple of important but still un(re)solved
problems.

$(i)$  {\it Factorization} of semi-inclusive processes:
an all-order factorization (in a covariant gauge) was studied in Ref.\
\cite{JMY04}, but definition (\ref{eq:tmd_v}) was used which contains
off-the-light-cone longitudinal gauge links.
An explicit proof of a factorization theorem with lightlike longitudinal
gauge links in the TMD PDFs is to our knowledge still lacking.

$(ii)$ {\it Relationship between TMDs and collinear PDFs}:
the generalized definition with lightlike longitudinal gauge links
(\ref{eq:uv_nr}) does indeed yield after integration an
$x-$\-de\-pen\-dent distribution function that obeys the DGLAP evolution
equation
\begin{equation}
  \int\! d^2 k_\perp \ f_1 (x, \mathbf{k_\perp}, \mu | n )
  =
  f_1 (x) \ , \  \mu \frac{d}{d\mu} \ f_1 (x, \mu)
  =
  {\cal K}_{\rm DGLAP} \otimes f_1 (x, \mu) \ .
\end{equation}
The reason is that the overlapping singularities in the real and the
virtual gluon contributions cancel against each other after carrying
out the $\mathbf{k_\perp}$-integration.
In contrast, for an axial but not-lightlike gauge, one gets a function
containing longitudinal gauge links off-the-light-cone along the vector
$v = (v^+, v^-, \mathbf{0_\perp})$, i.e.,
\begin{equation}
  \int\! d^2 k_\perp \ f_1 (x, \mathbf{k_\perp}, \mu | v )
  =
  f_1 (x, \mu | v ) \ \ , \  \ \mu \frac{d}{d\mu} \ f_1 (x, \mu | v )
  =
  {\cal K}_{v} \otimes f_1 (x, \mu | v ) \ \ , \ \
  {\cal K}_{v} \neq {\cal K}_{\rm DGLAP} \ .
\end{equation}
The RG-properties of this object differ from those of the TMD PDF
with lightlike longitudinal gauge links (which fulfills the DGLAP
equation shown above).
This difference leads, in particular, to different evolution equations
and different phenomenological implications.


\end{document}